\documentclass[3p,times]{elsarticle}

\usepackage{vendor/ecrc}
\usepackage[T1]{fontenc}
\usepackage[latin9]{inputenc}
\usepackage{amsmath}
\usepackage{amssymb}
\usepackage{booktabs}
\usepackage{enumerate}
\usepackage{enumitem}
\usepackage{pgfplots}
\usepackage[figuresright]{rotating}
\usepackage{subcaption}
\usepackage{xspace}

\usepackage{tikz}
\usepackage{tikzscale}

\usepackage{pgfplots}
\pgfplotsset{compat=1.8}
\usepgfplotslibrary{statistics}

\usepackage[toc, nowarn]{glossaries}
\makeglossaries
\newacronym{ai}{AI}{artificial intelligence}
\newacronym{gi}{GI}{gastrointestinal}
\newacronym{ml}{ML}{machine learning}
\newacronym{cnn}{CNN}{machine learning}
\newacronym{xai}{XAI}{explainable artificial intelligence}
\newacronym{shap}{SHAP}{SHapley Additive exPlanations}
\newacronym{gradcam}{GradCAM}{Gradient-weighted Class Activation Mapping}
\newacronym{lime}{LIME}{Local Interpretable Model-Agnostic Explanations}

\usepackage{hhline}
\usepackage{multirow}
\usepackage{makecell}

\volume{00}
\firstpage{1}

\journalname{Artificial Intelligence in Medicine}
\runauth{Hicks et al.}

\newcommand{\numparticipants}[0]{$54$\xspace}

\begin{document}

\begin{frontmatter}
\dochead{}
\title{Visual explanations for polyp detection: 
    How medical doctors \\ assess intrinsic versus extrinsic explanations}

\author[1]{Steven Hicks}
\author[1]{Andrea Stor\r{a}s}
\author[1,6]{Michael Riegler}
\author[1]{Cise Midoglu}
\author[1]{Malek Hammou}
\author[3]{Thomas de Lange}
\author[4]{Sravanthi Parasa}
\author[1,5]{P\r{a}l Halvorsen}
\author[2,1]{Inga Str{\" u}mke}

\address[1]{Department of Holistic Systems, SimulaMet, Norway}
\address[2]{Department of Computer Science, Norwegian University of Science and Technology, Norway}
\address[3]{Medical Department, Sahlgrenska University Hospital-M{\"o}lndal, Sweden}
\address[4]{Department of Gastroenterology, Swedish Medical Group, WA, USA}
\address[5]{Department of Computer Science, Oslo Metropolitan University, Norway}
\address[6]{Department of Computer Science, UIT The Artic University of Norway, Norway}


\begin{abstract}
Deep learning has in recent years achieved immense success in all areas of computer vision and has the potential of assisting medical doctors in analyzing visual content for disease and other abnormalities. However, the current state of deep learning is very much a black box, making medical professionals highly skeptical about integrating these methods into clinical practice. Several methods have been proposed in order to shine some light onto these black boxes, but there is no consensus on the opinion of the medical doctors that will consume these explanations. This paper presents a study asking medical doctors about their opinion of current state-of-the-art explainable artificial intelligence methods when applied to a gastrointestinal disease detection use case. We compare two different categories of explanation methods, intrinsic and extrinsic, and gauge their opinion of the current value of these explanations. The results indicate that intrinsic explanations are preferred and that explanation.
\end{abstract}

\begin{keyword}
Explainable Artificial Intelligence \sep Deep Learning \sep Medical Image Analysis
\end{keyword}
\end{frontmatter}

\section{Introduction}\label{section:introduction}
Deep learning is becoming an increasingly popular method for analyzing medical data to perform tasks like lesion detection or disease classification. However, despite the prevalent use of deep learning in medical research, deep learning is rarely deployed in a clinical setting~\cite{Kelly2019}. There are several factors that make using deep learning-based systems in medicine problematic, like the potential legal ramifications of incorrect diagnoses or the presence of unintended biases against a specific race or gender. Many of these issues stem from a general lack of explainability and interpretability in the employed deep learning algorithms. Deep neural networks are complex statistical models that consist of millions, if not billions, of parameters, making it difficult to understand what reasoning lies behind a specific prediction. \Gls{xai} aims to solve the issue of explainability and interpretability by providing methods that aim to explain the internal decision-process of the neural network in a more digestible and understandable manner. Several \gls{xai} methods have been proposed, where \gls{shap}~\cite{Lundberg2017SHAP} and salient-based explanations like \gls{gradcam}~\cite{Selvaraju:2017:ICCV.2017.74} are among the most popular techniques for image-based models. These methods provide an overlay that signifies what regions of the image contributed to the predicted output, making them relatively easy to understand among a non-technical audience. Several studies stress the importance of explanations that can be interrupted by non-tech-savvy users like medical doctors or clinicians for them to better understand the underlying reasoning behind a prediction~\cite{vellido2020XAIimportance}. However, there is no consensus on what explanation methods are preferred or if medical professionals actually find them useful. Similar studies have been done on the general population~\cite{NEURIPS2020_2c29d89c}, however, a study on the opinion of domain experts on \gls{xai} has yet to be done as far as we know.

This paper presents a study on gathering feedback from medical doctors regarding the current state-of-the-art \gls{xai} methods used to explain the prediction of computer-vision-based deep learning models. The study was conducted using automatic detection of colon polyps as a use-case, where a deep learning-based model is tasked with classifying images as either containing a polyp or not. Polyps are lesions within the bowel detectable as mucosal outgrows. Polyps are flat, elevated, or pedunculated and distinguished from normal mucosa by color and surface pattern. Most bowel polyps are harmless, but some have the potential to grow into cancer. Therefore, detection and removal of polyps are important to prevent the development of colorectal cancer. Since the doctors may overlook polyps, automatic detection would most likely improve examination quality. In live endoscopy, information about the endoscope configuration helps determine the current localization of the endoscope tip (and thereby also the polyp site) within the length of the bowel. Automatic computer-aided detection of polyps would be valuable for diagnosis, assessment, and reporting, and is currently a very popular research area in medical \gls{ai}~\cite{Hoerter2020,10.3389/fmed.2021.709347}. Due to its timeliness and clear objective, we find that the \gls{gi} use-case makes a perfect case-study for evaluating \gls{xai} explanation methods for medical use-cases. Please note that this study only looks at the explanation of 2-dimensional visual prediction models.

The rest of this paper is organized as follows: In Section~\ref{section:background}, we provide background information on the state-of-the-art explanation methods, explanations and how their current state in medical sciences and define our research questions. In Section~\ref{section:study-design}, we describe the process through which we have designed, implemented, and ran a subjective user study involving \numparticipants participants. In Section~\ref{section:results}, we present our results from the qualitative user study. In Section~\ref{section:discussion}, we discuss our findings and derive a number of generalizable insights regarding the applicability of \gls{xai} in medicine. We also put this in context with current work in the medical domain on XAI and how useful it is or not. There is in general a misconception of some people that argue XAI is not important or important but most of these works are rather personal opinions and not based on proper studies. We want to check how their claims compare to our findings. In Section~\ref{section:conclusion}, we conclude the paper. 

\section{Background and related work}\label{section:background}
\Gls{xai} is the sub-field of \gls{ai} dedicated to explaining \gls{ai} systems that are opaque or non-intuitive to humans. Different end-users have different needs for explanations, ranging from the developer who wishes to improve the system and ensure its robustness to the doctor using the system as decision support in the clinic and wanting to verify the veracity of the system's findings and potentially communicate this to the patient. Recent advances in the \gls{xai} literature almost exclusively concern methods designed to explain the behavior of complex \gls{ml} models. The \gls{xai} literature is large and rapidly developing, and we do not attempt to give an overview here. For the sake of simplicity, we compare two categories of \gls{xai} methods, intrinsic and extrinsic explanations. Intrinsic explanations cover the methods that aim to explain a model's internals through analyzing the model weights, which mostly includes saliency-based methods. Extrinsic explanations aim to explain the model using external input like \gls{shap} or \gls{lime}. In the following, we give a brief explanation of intrinsic and extrinsic explanation methods, and explain which explanation method we use to represent each category.

\subsection{Intrinsic explanations}\label{sec:gradcam}
As previously explained, intrinsic explanations aim to explain the predictions of a model by looking at the internal weights to provide some reasoning behind a specific output, and is the most common method for explaining deep neural networks. There is a large variety of such methods available, including~\cite{simonyan2014deep, Selvaraju:2017:ICCV.2017.74, zhang2017mdnet, smilkov2017smoothgrad, sundararajan2017axiomatic, bach2015pixel, kindermans2017learning, montavon2017explaining}, and it is not obvious which, if any, method is superior. In this study, we use \gls{gradcam}, which is arguably the most popular method for intrinsic-based visual explanations. Moreover, \gls{gradcam} has passed several sanity checks, as opposed to other popular intrinsic explanation methods~\cite{Adebayo2018SanityCheck}. \Gls{gradcam} highlights the important parts of the image for a predicted class based on the activated neurons in a specific layer of a neural network model. First, the gradients for the class are computed with respect to the feature maps of a layer in the model. The weights that are important for predicting a selected class are then obtained in order to identify what parts of the image contribute to the prediction, which can then be mapped back to the input. The heat maps follow the standard Jet color mapping, which consists of a gradient from red to green and green to blue, where red indicates the most important areas of the image, yellow indicates less important areas, and blue marks the least important areas. Examples from the study can be seen in row 2 of Figure~\ref{figure:case_examples}. The user selects which layer in the model to extract the heat maps from. Usually, later convolutional layers, i.e., layers that are close to the output layer, are helpful in order to highlight higher-level details. Moreover, the heat maps will depend on the model architecture since this will affect the activations for the selected layer. Consequently, many different heat maps can be generated for the same image. This means that some heat maps might be regarded as useful for the users, while others might not. An advantage of \gls{gradcam} is that it visually explains the inner workings of the neural network model, making it easier to understand.


\subsection{Extrinsic explanations}\label{sec:SHAP}
Extrinsic explanation methods, which also fall into the categories referred to as model-agnostic and model-independent, treat the model as a phenomenon and present information about its emergent behavior. In the case of image classification, the presentation is visually similar to the aforementioned class of methods, i.e., a heat map superimposed on the image, but involves occlusion~\cite{zeiler2014visualizing} and perturbation of image segments, for which there are significantly fewer methods of this kind available. The \gls{shap}~\cite{Lundberg2017SHAP} library is widely used for explaining \gls{ml} models, popular for its solid theoretical basis in game theory. The name of the \gls{shap} package~\cite{Lundberg2017SHAP} is an acronym for Shapley additive explanations, and the method is based on the Shapley decomposition, which is a solution concept from cooperative game theory. \Gls{shap} simulates feature absence by sampling from a background data set. Therefore, the resulting \gls{shap} value indicates how much the value of a feature causes the model's prediction to move away from the average prediction across the data.
For images, the systematic removal of all the pixels of which an image consists is infeasible, so \gls{shap} instead groups pixels based on their relative characteristics. For this, \gls{shap} uses an external computer vision-based segmentation algorithm~\cite{Achanta10slicsuperpixels}. The coarseness of the final \gls{shap} heat map is user-adjusted, and smaller pixel groups require more computation resources. To summarize, \gls{shap}, when applied to images, produces a heat map indicating which parts of the image support and oppose a classification relative to a chosen background data set. The built-in color scheme, also used in our study, uses shades of pink and blue superimposed on the image to indicate that a region supports (pink) or opposes (blue) the model prediction. Examples from the study can be seen in row 3 of Figure~\ref{figure:case_examples}.

\section{Study design}\label{section:study-design}

\newcommand{\lastcolumnwidth}[0]{7cm}

\begin{table*}
\small
\centering
\caption{The questions that were asked to the participants in the final feedback form. Please note that \textit{Explanation A} refers to the intrinsic explanation and \textit{Explanation B} is the extrinsic explanation.}
\label{table:feedback-form-questions}

\begin{tabular}{l l}

\toprule
Type & Question \\ 
\midrule
\multirow{9}{*}{Likert scale (1-10)} & Explanation (A) increased my understanding of the result. \\
& Explanation (B) increased my understanding of the result. \\
& Explanation (A) increased my trust in the AI model. \\ 
& Explanation (B) increased my trust in the AI model. \\ 
& I found the colors used to visualize explanation (A) to be appropriate. \\ 
& I found the colors used to visualize explanation (B) to be appropriate. \\ 
& Explanation (A) frequently highlighted the correct area in the image. \\
& Explanation (B) frequently highlighted the correct area in the image. \\
& It is important that an explanation accompanies a prediction. \\
\midrule
\multirow{3}{*}{Multiple choice} & Do you prefer to have an explanation, or would you rather only know the prediction? \\
& Which type of explanation would be useful in clinical practice? \\
& Would you prefer that explanations for the predictions be shown during or after the procedure? \\ 
\midrule
\multirow{2}{*}{Free form} & What do you think of explanation method (A)? \\
& What do you think of explanation method (B)? \\
\bottomrule




\end{tabular}
\end{table*}

The primary goal of this study was to quantify the value of current state-of-the-art \gls{xai} methods from the perspective of medical doctors. Over the course of five months (September 2021 - February 2022), we sent out a survey invitation to a number of different medical doctors located in different parts of the world together with a short video explaining the study\footnote{https://youtu.be/JJ8uc5gReko}. No compensation was given for taking part in the study. In this section, we describe the motivation and thought process behind building the study. This includes the development of the online survey, the implementation of the deep learning model used to generate the question cases, and the dataset used.

\subsection{Survey}\label{section:feedback-questions}

The survey was built using the open-source framework \textit{Huldra}, which is a framework for collecting crowdsourced feedback on multimedia assets~\footnote{https://github.com/simula/huldra}. The framework allows for the collection of participant responses in a storage bucket hosted on the cloud, from where they can be retrieved in real-time by survey organizers, using credentials, immediately after the first interaction of each participant. The survey consisted primarily of four distinct parts; registration, orientation, case questionnaire, and feedback.

The first part of the survey asked participants to register with their name (optional), email (optional), country, academic degree(s), their field of expertise, and how many years they have been active in the field. The second part oriented the participants on what they could expect from the survey and gave some background information on the two explanation methods that would be compared during the survey. The main part of the survey consisted of 10 cases where a model predicted that an image from the \gls{gi} tract contained a polyp or not. The prediction was shown together with the image, alongside the two explanation methods that support the prediction. Here, the participants were asked to select which of the two explanation methods they found most helpful. The cases were shuffled on a per-participant basis, meaning the order of which the cases were shown was not the same between two participants. As the attention span may differ between the first and last case, we wanted to avoid any bias that could be introduced through the ordering of the different cases. The last part of the survey contained a feedback form consisting of 14 questions (see Table~\ref{table:feedback-form-questions}) that were meant to derive a summary of the doctor's overall perception of the two explanation methods. The participants were also shown a summary of their previous answers, where they had the option to go back and review/change the selection for specific cases.

{

\newcommand{\figurewidth}{0.19\textwidth}

\begin{figure}
  \centering
     \begin{subfigure}[b]{\figurewidth}
         \centering
         \includegraphics[height=\textwidth, width=\textwidth]{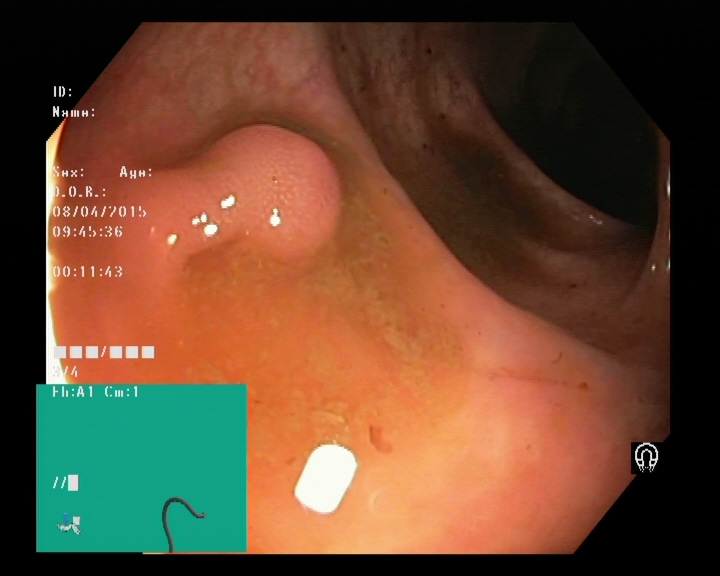}
         \caption{}
         \label{figure:user-study-cases-0b4f0370-b25b-4e6a-a7ac-b09097bfecc3-image}
     \end{subfigure}
          \begin{subfigure}[b]{\figurewidth}
         \centering
         \includegraphics[height=\textwidth, width=\textwidth]{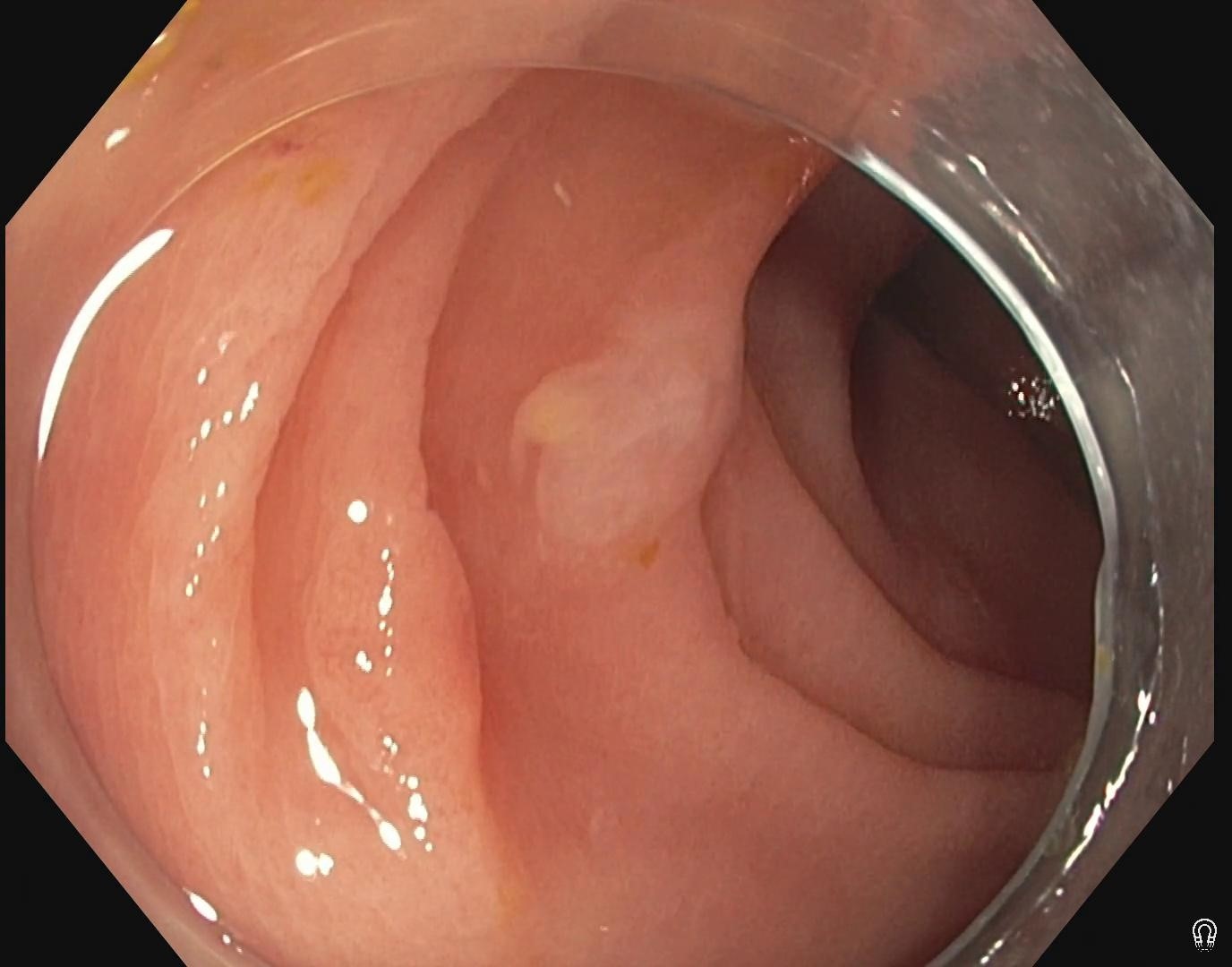}
         \caption{}
         \label{figure:user-study-cases-4_endocv2021_positive_34-image}
     \end{subfigure}
     \begin{subfigure}[b]{\figurewidth}
         \centering
         \includegraphics[height=\textwidth, width=\textwidth]{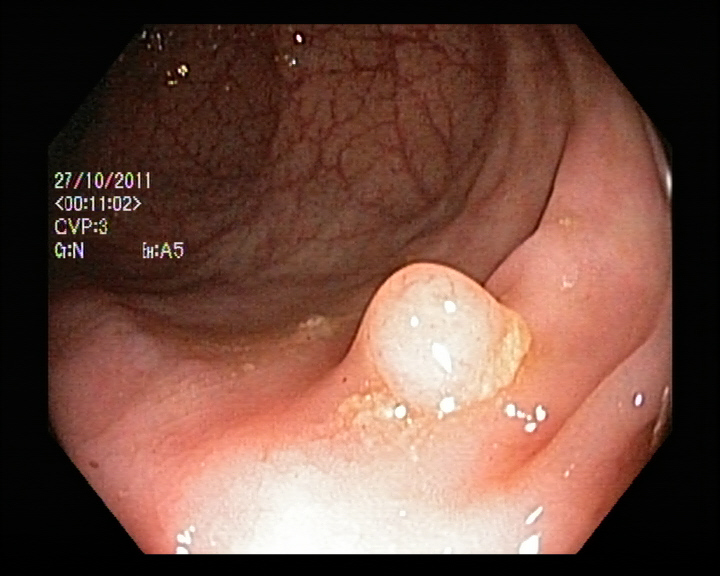}
         \caption{}
         \label{figure:user-study-cases-1dc7852b4443-image}
     \end{subfigure}
      \begin{subfigure}[b]{\figurewidth}
         \centering
         \includegraphics[height=\textwidth, width=\textwidth]{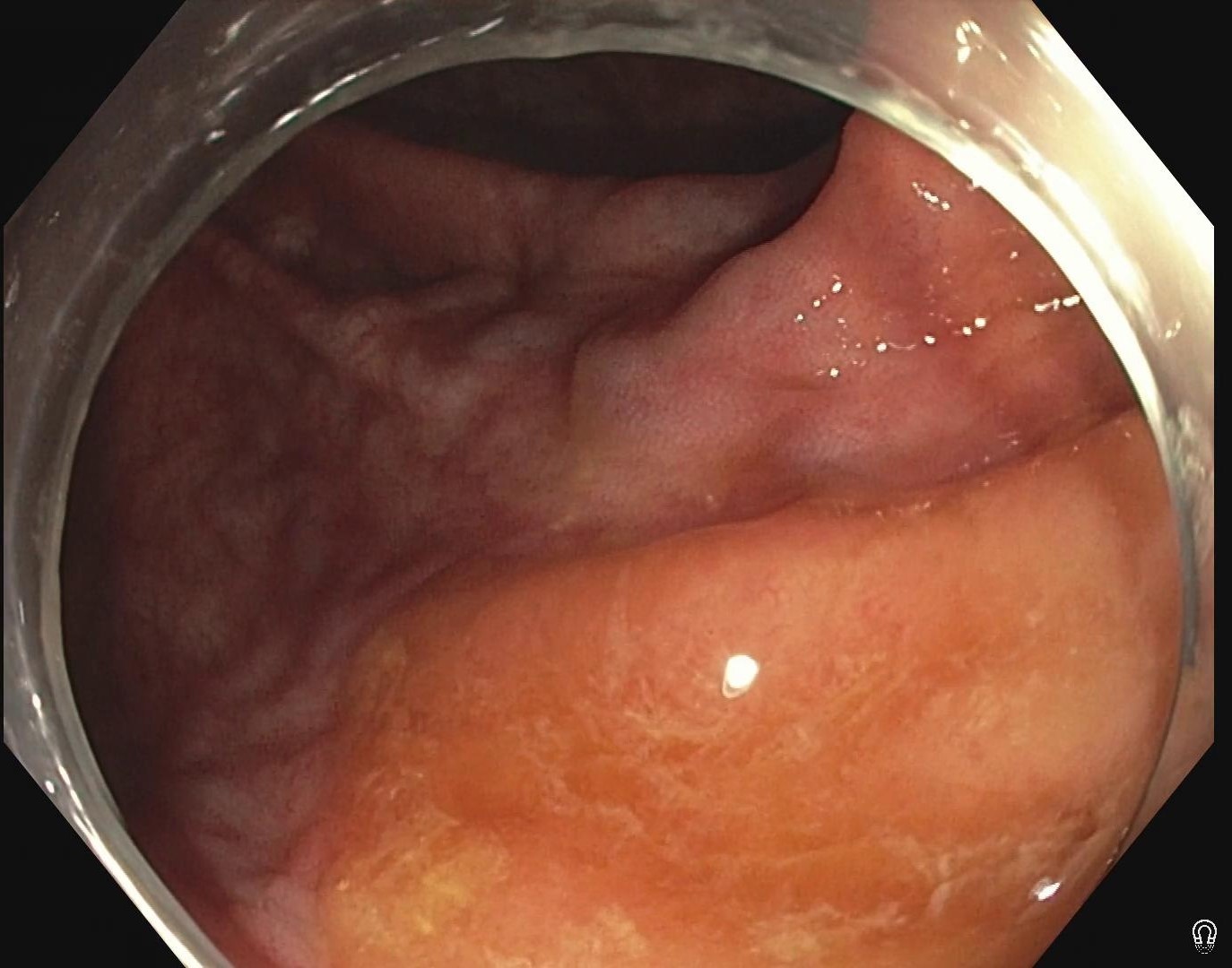}
         \caption{}
         \label{figure:user-study-cases-14_endocv2021_positive_902-image}
     \end{subfigure}
      \begin{subfigure}[b]{\figurewidth}
         \centering
         \includegraphics[height=\textwidth, width=\textwidth]{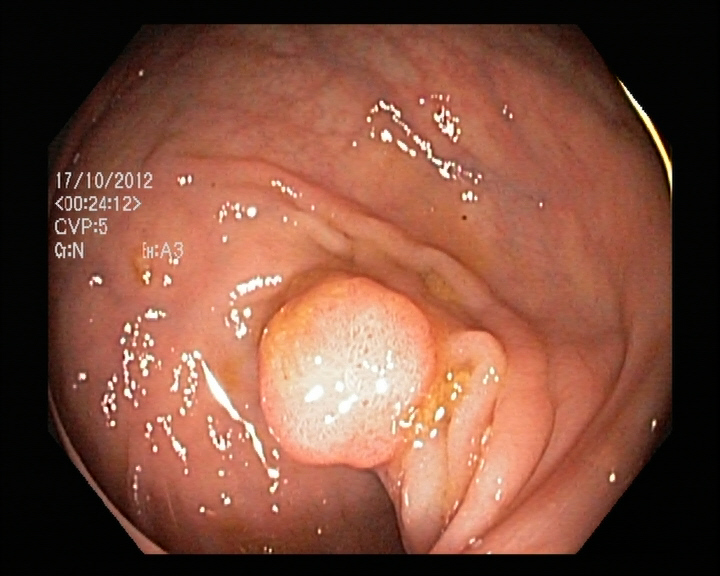}
         \caption{}
         \label{figure:user-study-cases-2f19c3270595-image}
     \end{subfigure}

     \begin{subfigure}[b]{\figurewidth}
         \centering
         \includegraphics[height=\textwidth, width=\textwidth]{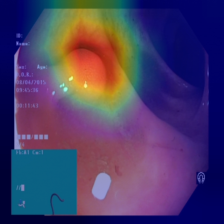}
         \caption{}
         \label{figure:user-study-cases-b09097bfecc3-a}
     \end{subfigure}
          \begin{subfigure}[b]{\figurewidth}
         \centering
         \includegraphics[height=\textwidth, width=\textwidth]{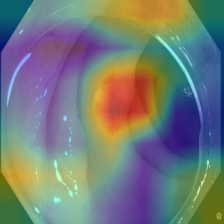}
         \caption{}
         \label{figure:user-study-cases-4_endocv2021_positive_34-a}
     \end{subfigure}
     \begin{subfigure}[b]{\figurewidth}
         \centering
         \includegraphics[height=\textwidth, width=\textwidth]{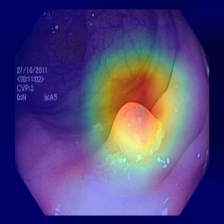}
         \caption{}
         \label{figure:user-study-cases-1dc7852b4443-a}
     \end{subfigure}
      \begin{subfigure}[b]{\figurewidth}
         \centering
         \includegraphics[height=\textwidth, width=\textwidth]{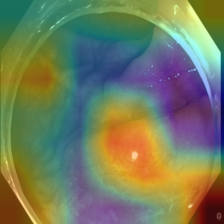}
         \caption{}
         \label{figure:user-study-cases-14_endocv2021_positive_902-a}
     \end{subfigure}
      \begin{subfigure}[b]{\figurewidth}
         \centering
         \includegraphics[height=\textwidth, width=\textwidth]{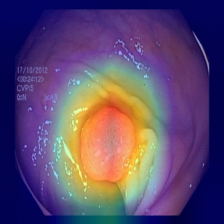}
         \caption{}
         \label{figure:user-study-cases-2f19c3270595-a}
     \end{subfigure}
     
     \begin{subfigure}[b]{\figurewidth}
         \centering
         \includegraphics[height=\textwidth, width=\textwidth]{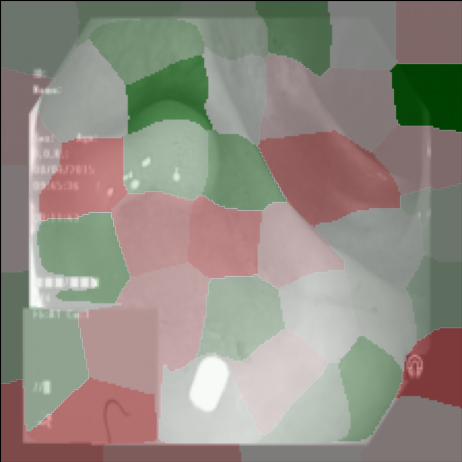}
         \caption{}
         \label{figure:user-study-cases-b09097bfecc3-b}
     \end{subfigure}
          \begin{subfigure}[b]{\figurewidth}
         \centering
         \includegraphics[height=\textwidth, width=\textwidth]{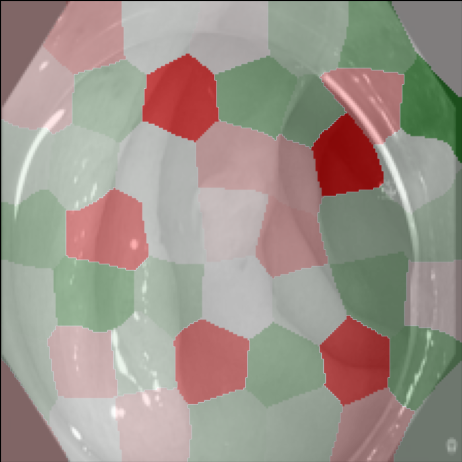}
         \caption{}
         \label{figure:user-study-cases-4_endocv2021_positive_34-b}
     \end{subfigure}
     \begin{subfigure}[b]{\figurewidth}
         \centering
         \includegraphics[height=\textwidth, width=\textwidth]{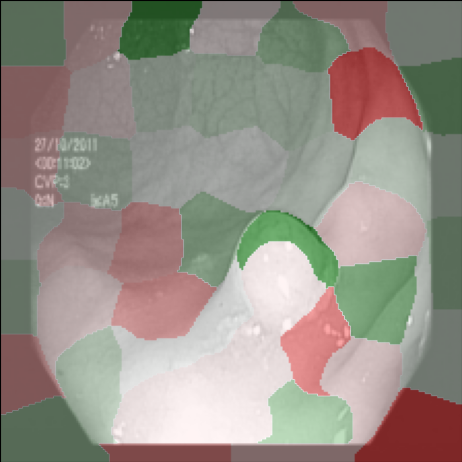}
         \caption{}
         \label{figure:user-study-cases-1dc7852b4443-b}
     \end{subfigure}
      \begin{subfigure}[b]{\figurewidth}
         \centering
         \includegraphics[height=\textwidth, width=\textwidth]{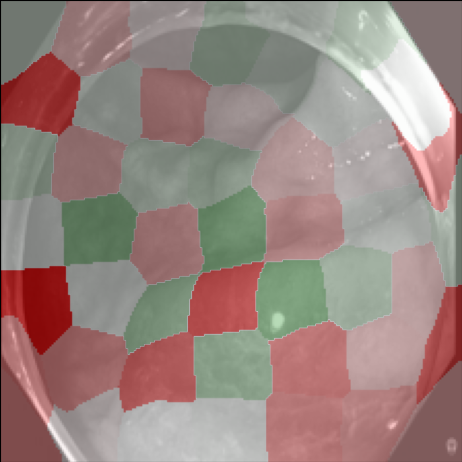}
         \caption{}
         \label{figure:user-study-cases-14_endocv2021_positive_902-b}
     \end{subfigure}
      \begin{subfigure}[b]{\figurewidth}
         \centering
         \includegraphics[height=\textwidth, width=\textwidth]{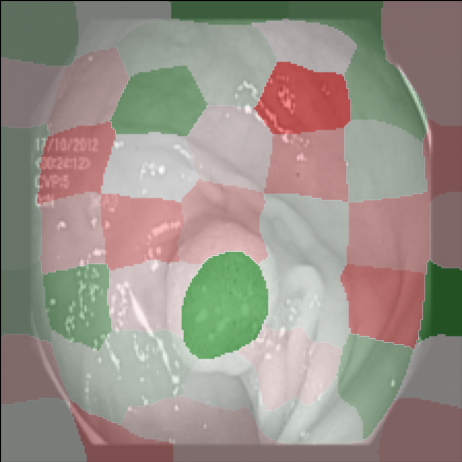}
         \caption{}
         \label{figure:user-study-cases-2f19c3270595-b}
     \end{subfigure}
     \caption{Five cases taken from the survey presented to each participant. The top row shows the image that was passed through the model to generate a prediction. The second row shows the intrinsic explanation method used to explain the prediction. The last row shows the extrinsic explanation method used to explain the prediction.}\label{figure:case_examples}
\end{figure}

}

\subsection{Dataset}\label{section:dataset}
The dataset used to sample the case images and train the deep neural network was \textit{Kvasir}~\cite{10.1145/3193289}, which is an open \gls{gi} dataset consisting of different findings from the \gls{gi} tract. \textit{Kvasir} consists of images annotated and verified by medical doctors (experienced endoscopists), including several classes showing anatomical landmarks, pathological findings, or endoscopic procedures in the GI tract, with hundreds of images for each class. The anatomical landmarks include Z-line, pylorus, cecum, etc., while the pathological finding includes esophagitis, polyps, ulcerative colitis, etc. In addition, several sets of images related to the removal of lesions are also provided, like \textit{dyed and lifted polyp} and \textit{dyed resection margins}. The dataset contains images with different resolutions from $720\times 576$ up to $1920\times 1072$ pixels. Some of the included classes of images have a green box in the lower-left corner that illustrates the position and configuration of the endoscope inside the bowel, using an electromagnetic imaging system (ScopeGuide, Olympus Europe). Examples from the dataset can be seen in row 1 of Figure~\ref{figure:case_examples}.

\subsection{Implementation of explanation methods}\label{section:implementation}
The model used to classify the images and generate the explanations was a \gls{cnn} based on the ResNet~\cite{He2015} architecture implemented in PyTorch and trained on a modified version of the aforementioned Kvasir~\cite{10.1145/3193289} dataset. The dataset was modified accommodate the use-case of distinguishing between images containing polyps and images of clean colon. As for the explanation methods, we used Captum~\cite{kokhlikyan2020captum} provided by PyTorch for the extrinsic explanations and an open implementation of GradCAM\footnote{https://github.com/jacobgil/pytorch-grad-cam} for the intrinsic explanations. The model was trained on what can be considered consumer-grade hardware, containing a Nvidia RTX 3090 GPU and an Intel i9 processor. The source code and more details on the implementation of the model used to generate the explanations can be found in our GitHub repository\footnote{https://github.com/simula/gi-xai-survey}.

\section{Survey results}\label{section:results}

{

\newcommand{\figurewidth}{0.33\textwidth}

\begin{figure}
  \centering
     \begin{subfigure}[b]{\figurewidth}
         \centering
         \includegraphics[height=\textwidth, width=\textwidth]{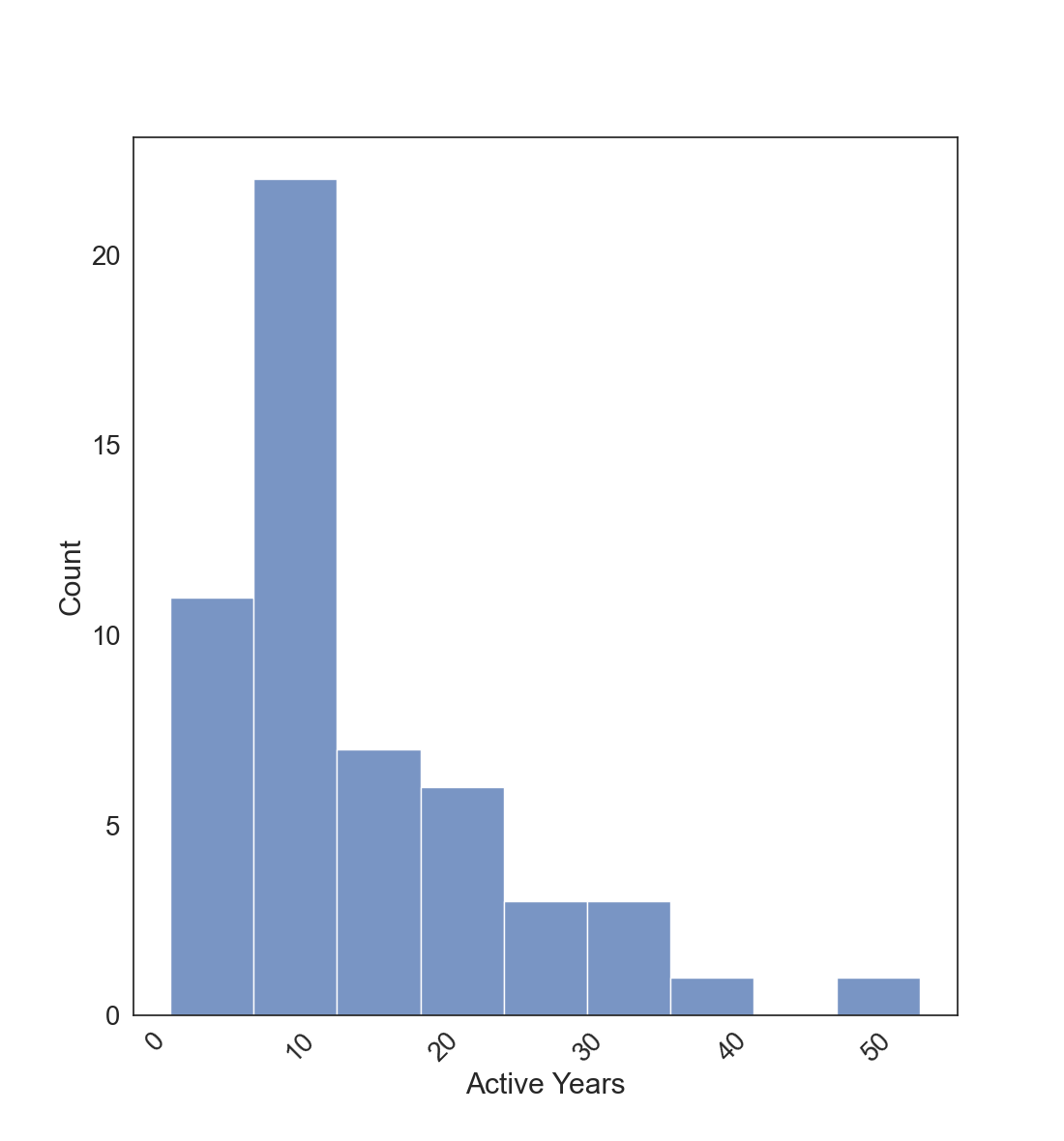}
         \caption{Number of active years in the medical field.}
         \label{figure:plot-active-years}
     \end{subfigure}
          \begin{subfigure}[b]{\figurewidth}
         \centering
         \includegraphics[height=\textwidth, width=\textwidth]{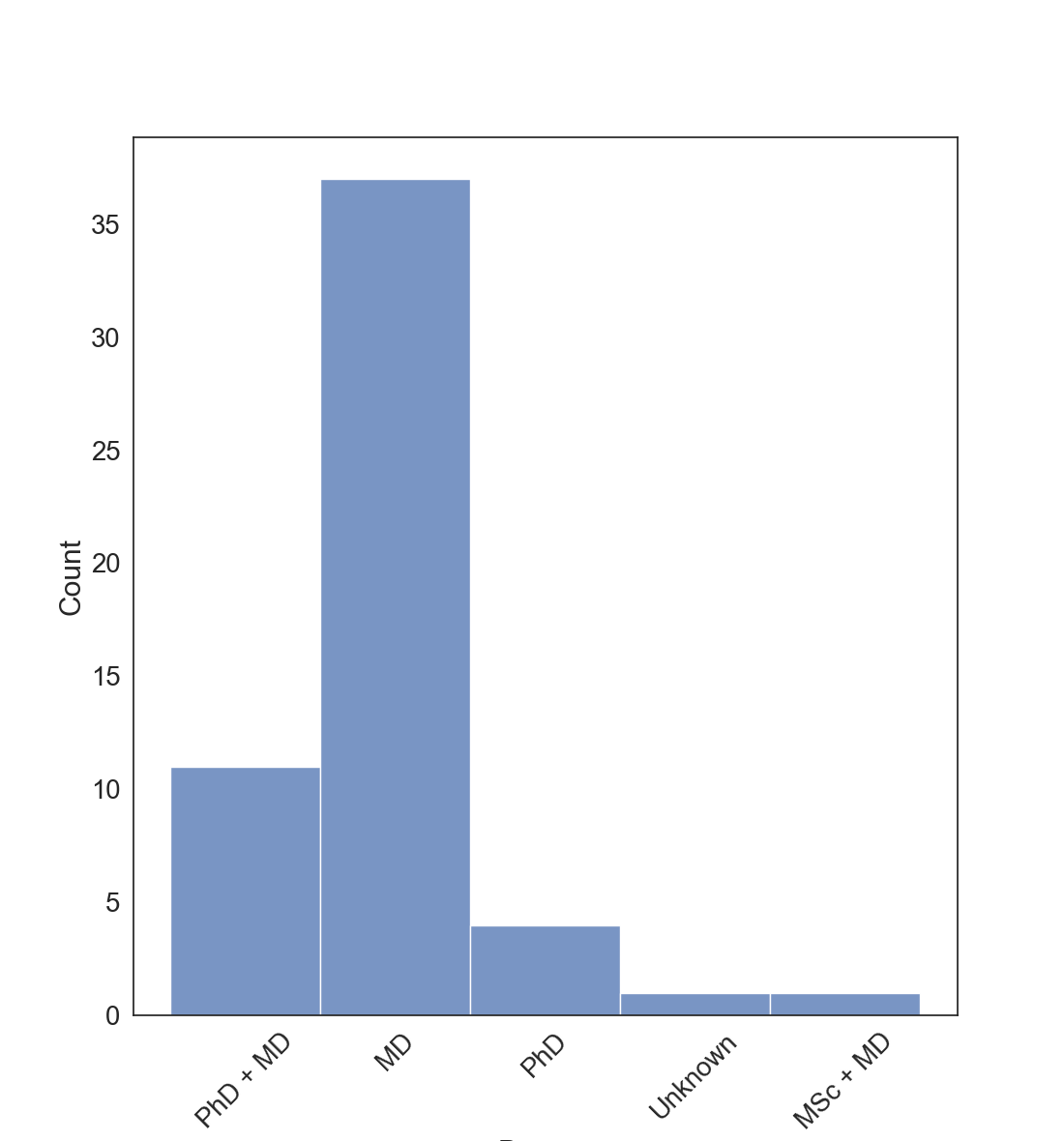}
         \caption{The degree(s) obtained by the participants.}
         \label{figure:plot-degree}
     \end{subfigure}
     \begin{subfigure}[b]{\figurewidth}
         \centering
         \includegraphics[height=\textwidth, width=\textwidth]{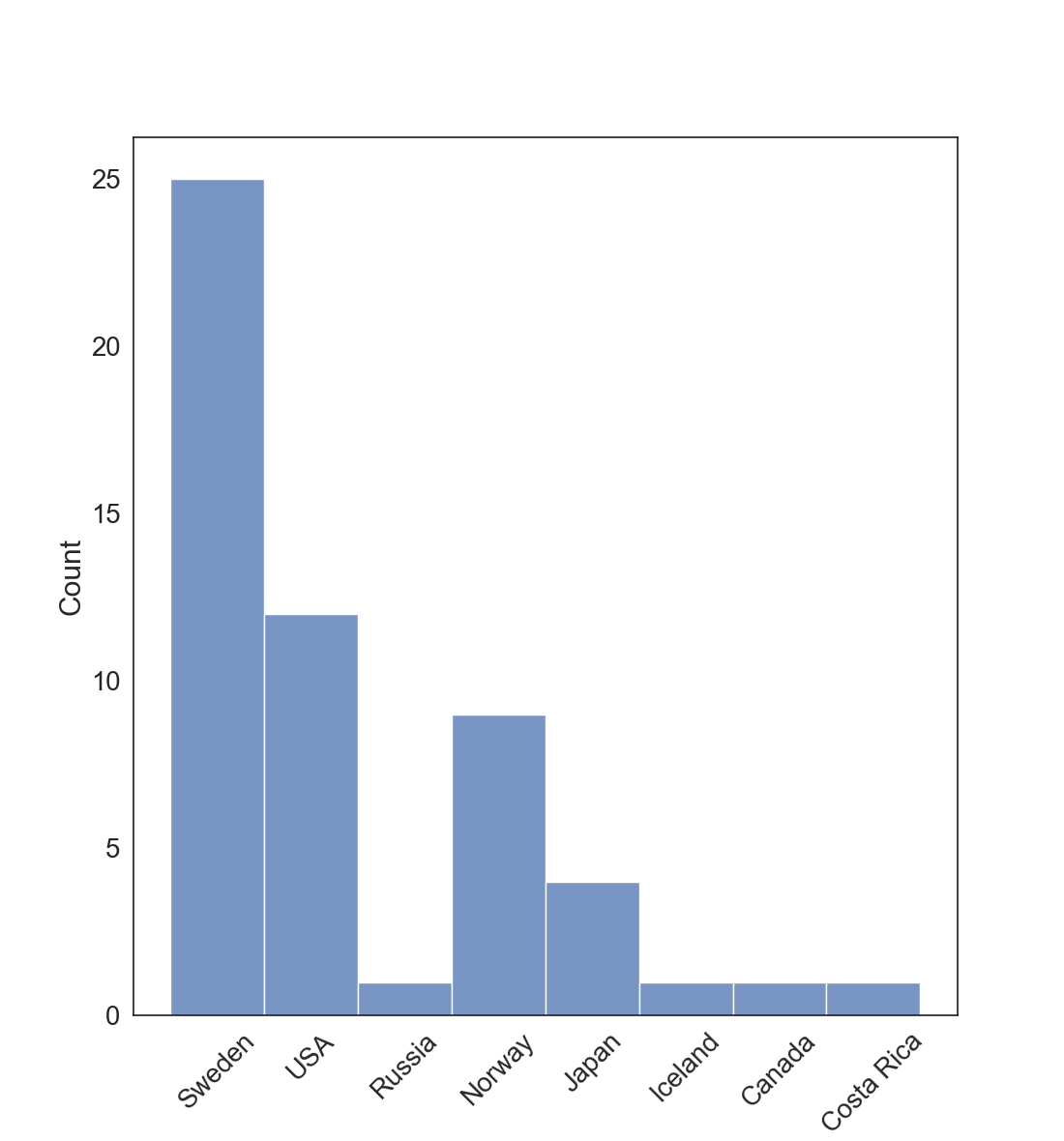}
         \caption{The country that the participant comes from.}
         \label{figure:plot-country}
     \end{subfigure}
     \caption{Plots presenting some statistics about the participants included in the study.}\label{figure:participant-statistics}
\end{figure}

}

The survey collected a total of 57 responses. Of these, \numparticipants were used in the final analysis. Among the initial responses were a few non-medical workers, including \gls{ai} specialists and marketers. As the primary motivation behind this study was to better understand the opinion of medical professionals working with \gls{ai}, we decided to filter out these and only keep the responses of the participants working in the medical field. Apart from non-medical participants, we also filtered out any incomplete submissions. In the end, the remaining participants came from eight different countries with a varying amount of experience in the medical field, ranging from just a few years to over 50 years. Figure~\ref{figure:participant-statistics} shows some plots regarding the participants statistics in terms of active years in the field, obtained degree(s), and the country that the participants come from. The rest of this section is organized by question category, where we present a summary of the participants' responses for the explanation case questions, Likert questions, multiple choice questions, and free-form questions.

\subsection{Explanation case responses}

\begin{table}
    \centering
    \caption{}\label{tab:intraclass_correlation_coefficient}
    \begin{tabular}{c c c c c c c c c}
         & & \multicolumn{2}{c}{95\% Confidence Interval} &  & \multicolumn{4}{c}{F Test with True Value 0} \\ \cmidrule{3-4} \cmidrule{6-9}
         & Interclass Correlation & Lower Bound & Upper Bound &  & Value & df1 & df2 & Sig \\
        \midrule
            Single Measures &  $0.067$ &  $0.023$ &  $0.220$ & &  $4.852$ & $9$ & $477$ & < $0.000$ \\
            Average Measures &  $0.794$ &  $0.559$ &  $0.938$ & &  $4.852$ & $9$ & $477$ & < $0.000$ \\
        \bottomrule
    \end{tabular}
\end{table}
\begin{table}
    \centering
    \caption{}\label{tab:overall_agreement}
    \begin{tabular}{c c c c c c c c}
         & & \multicolumn{3}{c}{Asymptotic}&  & \multicolumn{2}{c}{Asymptotic 95\% Confidence Interval} \\ \cmidrule{3-5} \cmidrule{7-8}
         & Kappa & Standard Error & z & Sig & & Lower Bound & Upper Bound \\
        \midrule
            Overall Agreement & $0.049$ & $0.008$ & $5.878$ & $<.001$ & & $0.033$ & $0.066$ \\
        \bottomrule
    \end{tabular}
\end{table}

To get a better understanding of the agreement between the different participants, we performed an inter-rater reliability test that was calculated for all explanation case responses (see Table~\ref{table:feedback-form-questions}) in order to measure the level of agreement between the study participants. Intra-class correlation (ICC) is one of the most common ways to investigate inter-rater reliability for ordinal variables~\cite{hallgren2012IRR}. An ICC value of $1$ corresponds to perfect agreement. Values between $0.70$ and $0.79$ are regarded as fair, values between $0.80$ and $0.89$ are good, while values of $0.90$ and above are excellent with respect to clinical relevance~\cite{cicchetti1994guidelines}. One of the strengths of ICC is that it takes the magnitude of disagreement between the raters into account, meaning that large disagreements result in lower ICC values than small disagreements~\cite{cicchetti1994guidelines}. The ICC was calculated for all the explanation cases in order to assess the level of agreement between the answers from the study participants. From Table~\ref{tab:intraclass_correlation_coefficient}, we see that the average measures ICC is $0.794 [0.559, 0.938]$, which means that the agreement is fair. The Fleiss' kappa value in Table~\ref{tab:overall_agreement} is, however, $0.049$. This corresponds to poor agreement between the study participants. Note that the agreement metrics reflect the agreement regarding the intrinsic and extrinsic explanation methods. Poor agreement in this context means that the study participants do not prefer the same explanation method. 

\subsection{Likert scale responses}

{

\newcommand{\figurewidth}{0.33\textwidth}

\begin{figure}
  \centering
     \begin{subfigure}[b]{\figurewidth}
         \centering
         \includegraphics[width=\textwidth]{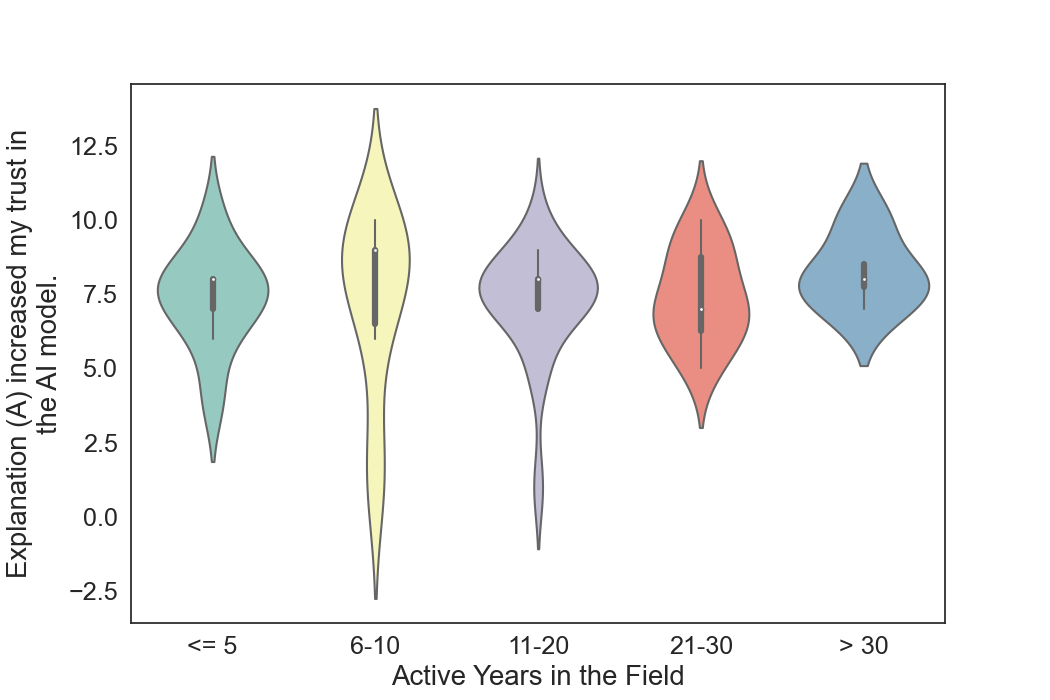}
         \caption{}
         \label{subfigure:trust-a}
     \end{subfigure}
          \begin{subfigure}[b]{\figurewidth}
         \centering
         \includegraphics[width=\textwidth]{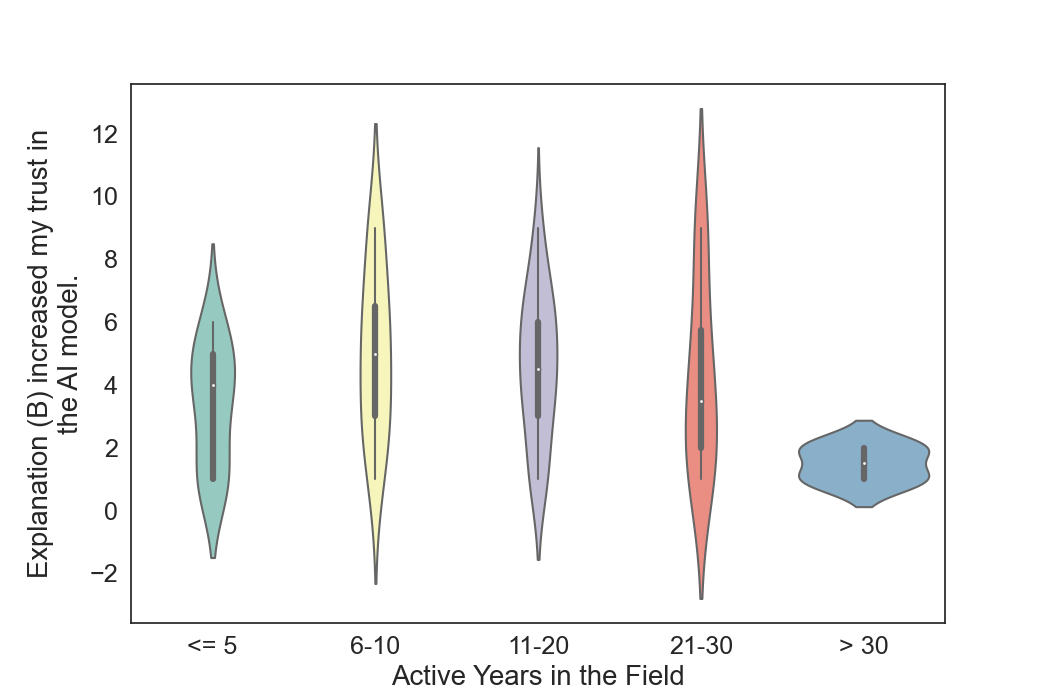}
         \caption{}
         \label{subfigure:trust-b}
     \end{subfigure}
     \begin{subfigure}[b]{\figurewidth}
         \centering
         \includegraphics[width=\textwidth]{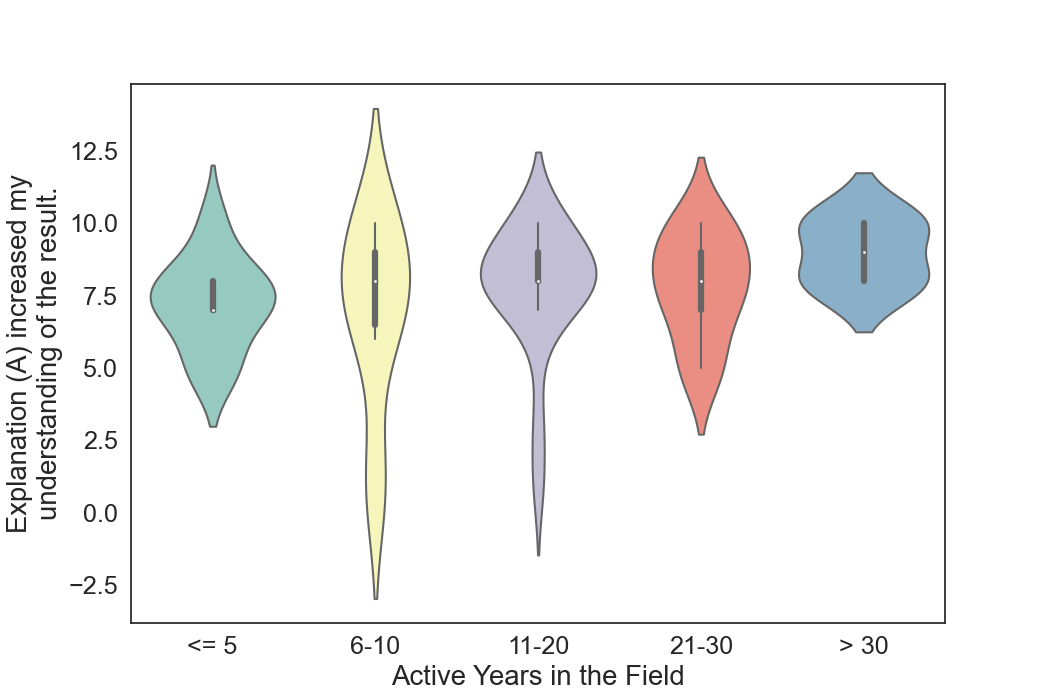}
         \caption{}
         \label{subfigure:understanding-a}
     \end{subfigure}
     
     \begin{subfigure}[b]{\figurewidth}
         \centering
         \includegraphics[width=\textwidth]{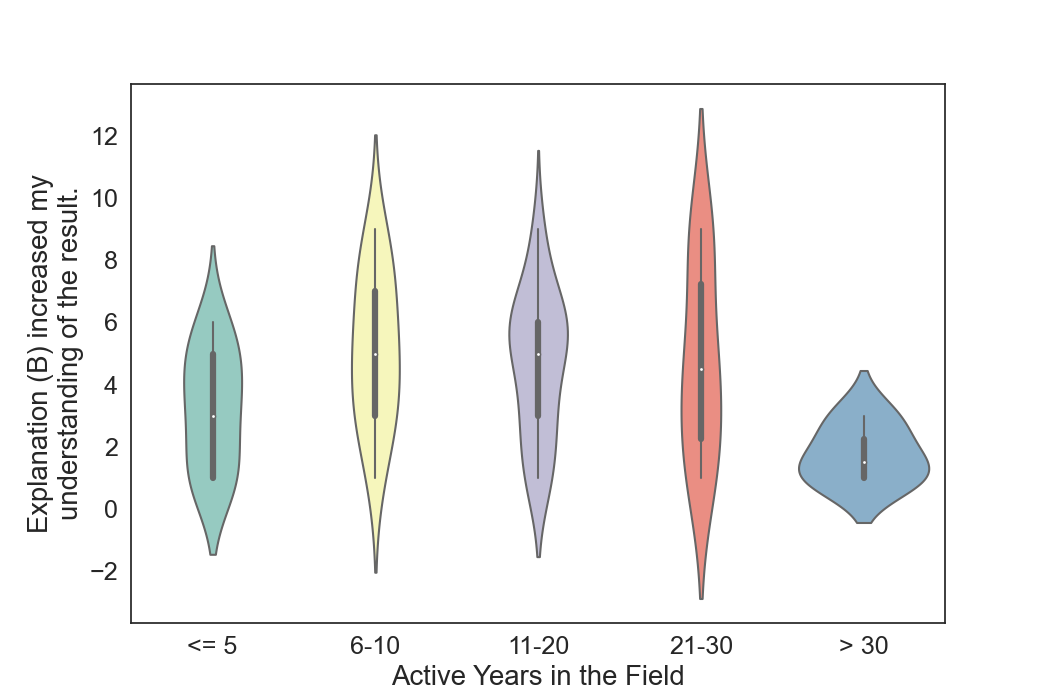}
         \caption{}
         \label{subfigure:understanding-b}
     \end{subfigure}
      \begin{subfigure}[b]{\figurewidth}
         \centering
         \includegraphics[width=\textwidth]{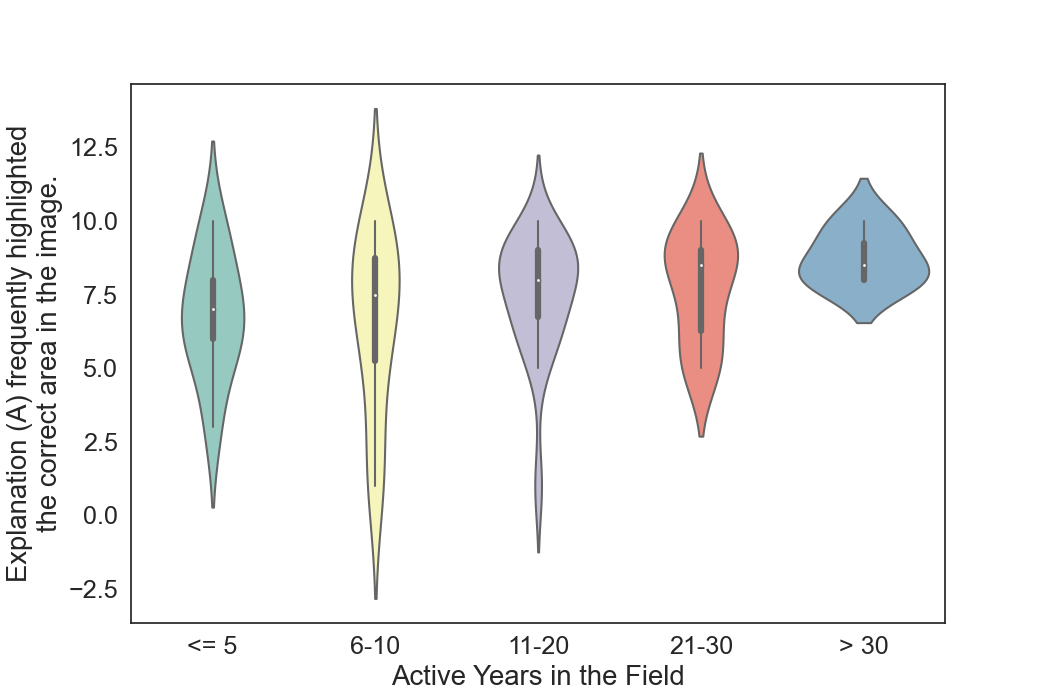}
         \caption{}
         \label{subfigure:spatial-a}
     \end{subfigure}
          \begin{subfigure}[b]{\figurewidth}
         \centering
         \includegraphics[width=\textwidth]{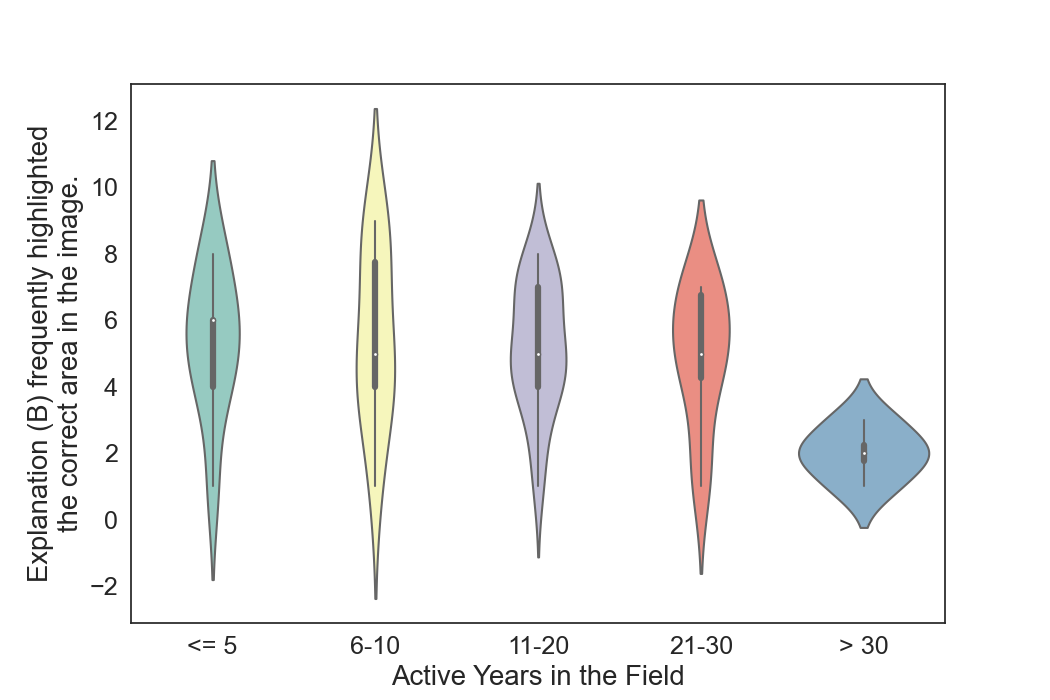}
         \caption{}
         \label{subfigure:spatial-b}
     \end{subfigure}
     
     \begin{subfigure}[b]{\figurewidth}
         \centering
         \includegraphics[width=\textwidth]{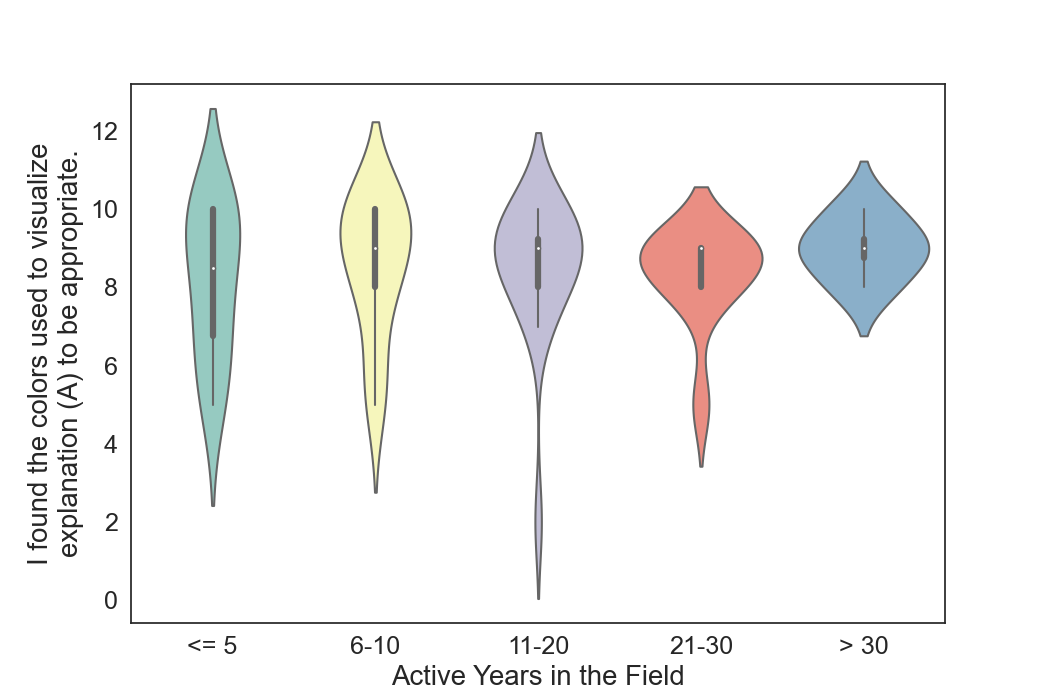}
         \caption{}
         \label{subfigure:colors-a}
     \end{subfigure}
     \begin{subfigure}[b]{\figurewidth}
         \centering
         \includegraphics[width=\textwidth]{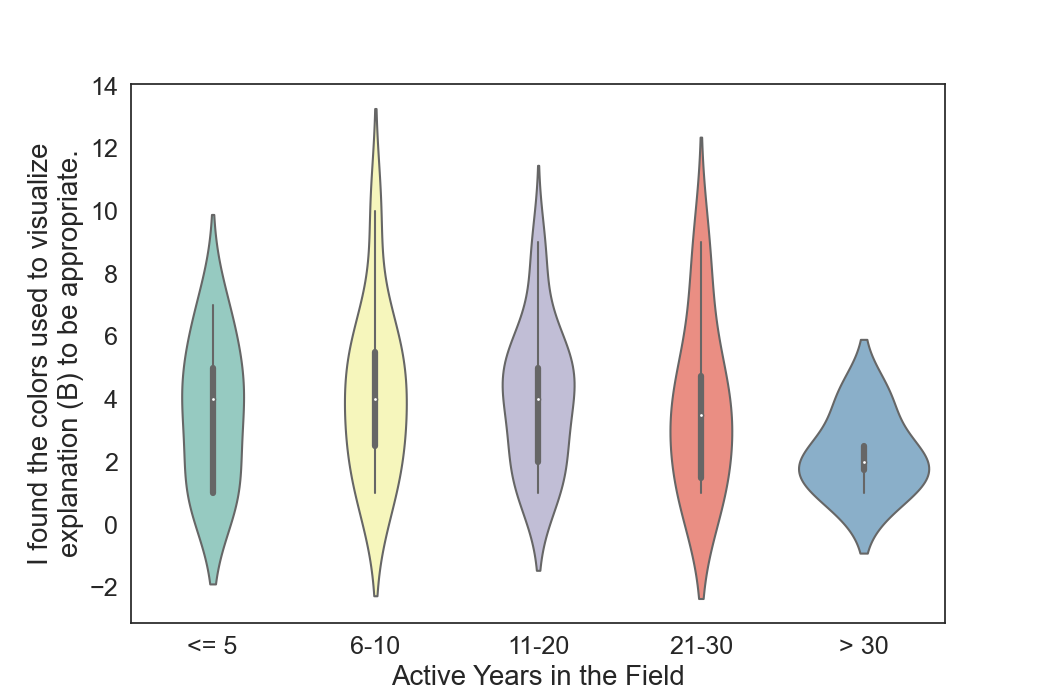}
         \caption{}
         \label{subfigure:colors-b}
     \end{subfigure}
     \begin{subfigure}[b]{\figurewidth}
         \centering
         \includegraphics[width=\textwidth]{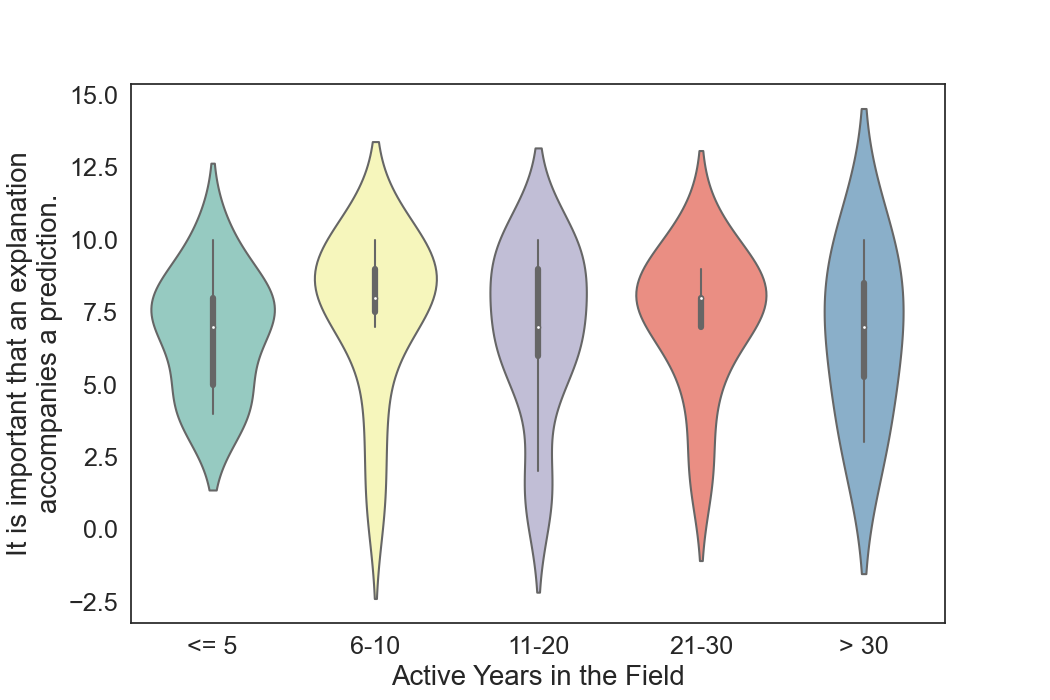}
         \caption{}
         \label{subfigure:importance}
     \end{subfigure}
     \caption{A collection of violin plots that presents an overview of the responses collected from the Likert questions. The answers are grouped by the number of years the person has been active in the medical field.}\label{figure:user-study-likert-violin-plots-experience}
\end{figure}

}

Figure~\ref{figure:user-study-likert-violin-plots-experience} shows a collection of violin plots showcasing the answers collected from the Likert questions asked in the survey. Comparing the plots asking about increasing trust in the model for each respective method (Figures~\ref{subfigure:trust-a} and~\ref{subfigure:trust-b}), we see that there generally seems to be more agreement that the intrinsic method induces more trust in the underlying model across all experience groups. This pattern continues when comparing the plots for understanding (Figures~\ref{subfigure:understanding-a} and~\ref{subfigure:understanding-b}), spatial relevance (Figures~\ref{subfigure:spatial-a} and~\ref{subfigure:spatial-b}), and color choices (Figures~\ref{subfigure:colors-a} and~\ref{subfigure:colors-b}).

\subsection{Multiple-choice responses}

{

\newcommand{\figurewidth}{0.33\textwidth}

\begin{figure}[ht!]
  \centering
     \begin{subfigure}[b]{\figurewidth}
         \centering
         \includegraphics[width=\textwidth]{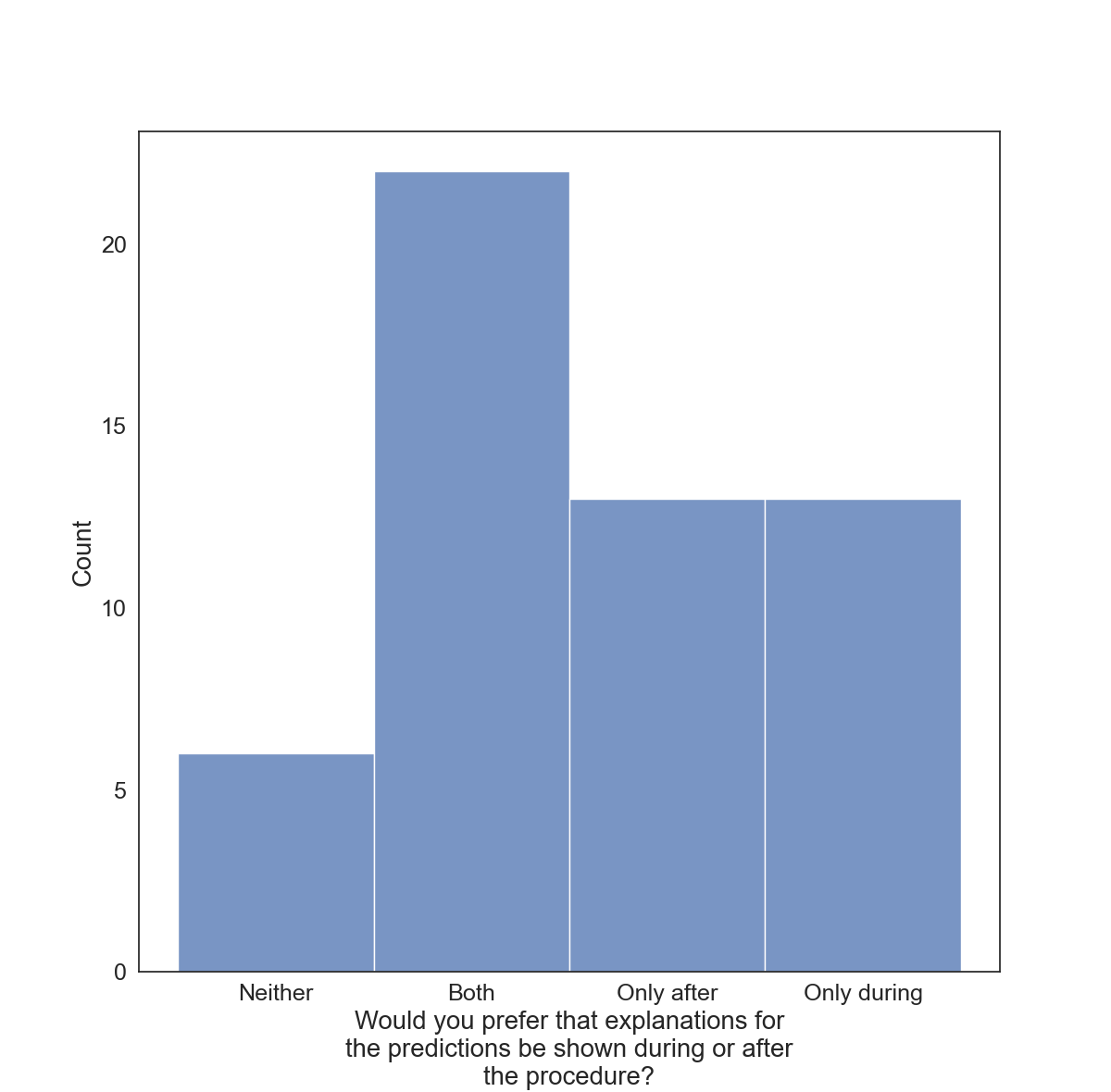}
         \caption{}
         \label{figure:histogram-before-after}
     \end{subfigure}
          \begin{subfigure}[b]{\figurewidth}
         \centering
         \includegraphics[width=\textwidth]{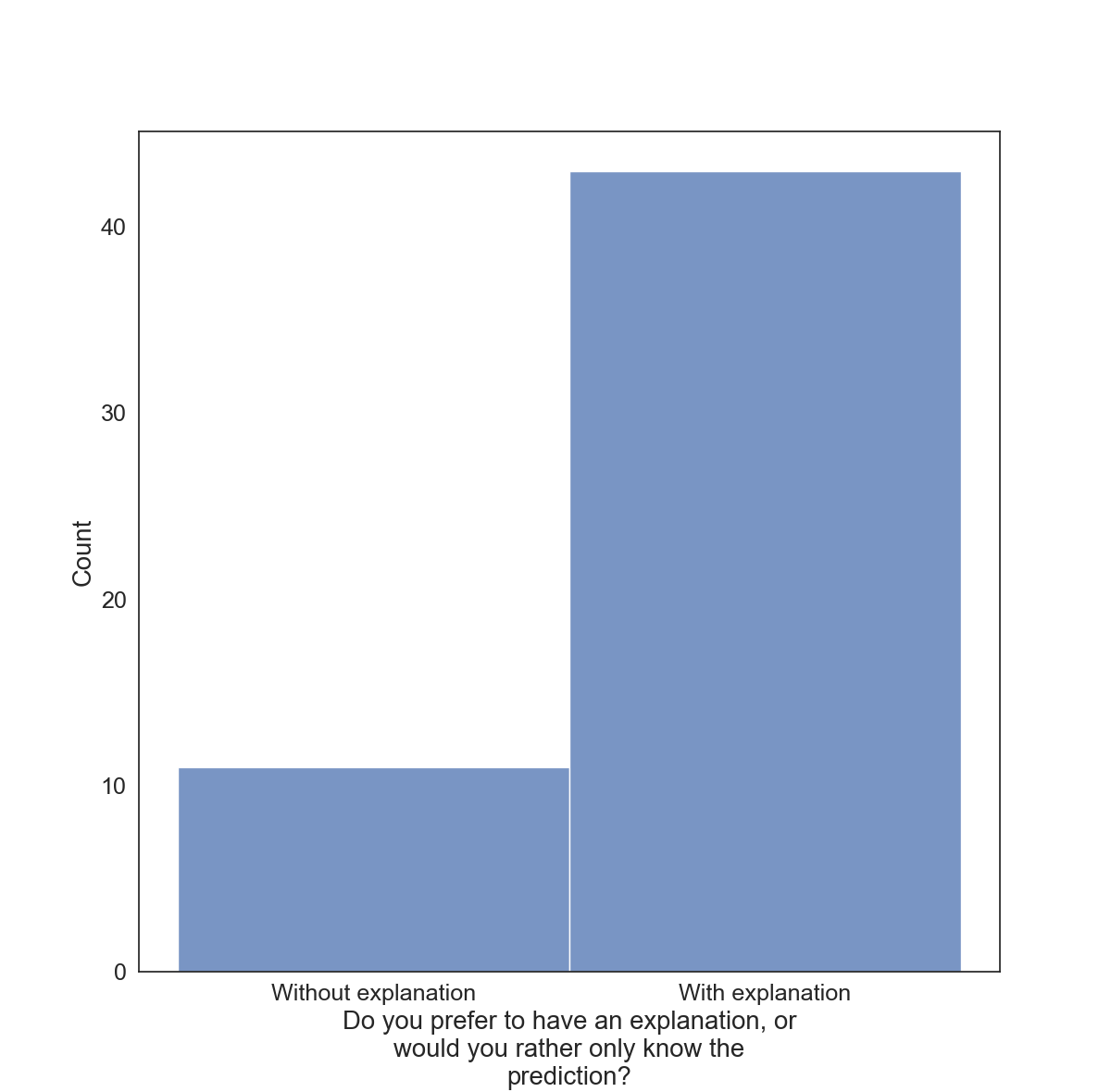}
         \caption{}
         \label{figure:histogram-explanation-or-not}
     \end{subfigure}
     \begin{subfigure}[b]{\figurewidth}
         \centering
         \includegraphics[width=\textwidth]{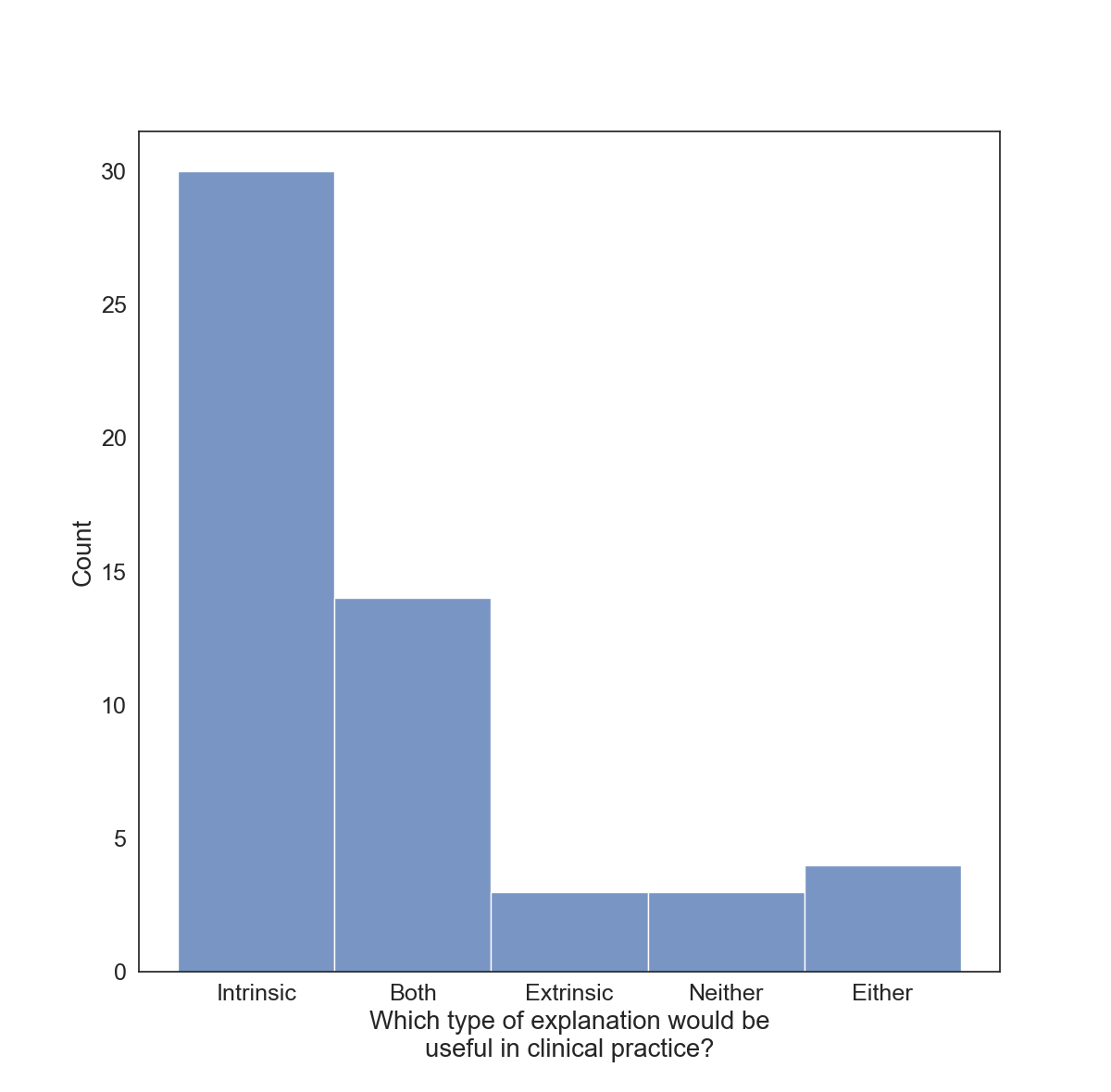}
         \caption{}
         \label{figure:histogram-most-useful}
     \end{subfigure}
     \caption{The responses collected from the multiple-choice questions asked in the survey.}
\end{figure}

}

According to the histogram plots in Figure~\ref{figure:histogram-before-after}, the majority of participants preferred to see model explanations both during and after the procedure, and explanations were preferred to no explanations, see Figure~\ref{figure:histogram-explanation-or-not}. From Figure~\ref{figure:histogram-most-useful}, we see that $30$ participants answered that the intrinsic explanation method would be useful in clinical practice, $14$ participants wanted both explanation methods, while $3$ wanted the extrinsic explanation method. $3$ participants answered that none of the explanation methods would be useful, and $4$ of them answered that either one or the other method would be useful.

\subsection{Free-form responses}
The survey included three questions with free-form responses, i.e., parts of the form where the participants could form their responses and elaborate freely. The three questions were
\begin{itemize}
    \item Do you prefer to have an explanation, or would you rather only know the prediction?
    \item What do you think of explanation method A?
    \item What do you think of explanation method B?
\end{itemize}
and we provide a summary of the responses to each question below.

\subsubsection{Do you prefer to have an explanation, or would you rather only know the prediction?}
One participant stated regarding explainability as a ``reasonable and popular expectation of AI systems'', but highlighted as a challenge that we humans upon being presented an explanation ``make human interpretations/assumptions around what the explanation indicates about the underlying AI process [although] these interpretations may not actually be an accurate reflection of what is actually happening'', along with a reference to~\cite{GHASSEMI2021e745}.

Another participant pointed out that ``in clinical practice we do not have enough time to check the explanation during colonoscopy.''. A third contrasted clinical work to research, stating that they prefer no explanation during the former but receiving an explanation during the latter.

\subsubsection{What do you think of explanation method A?}
As indicated by the Likert scale results, most respondents prefer the intrinsic explanation method, which was reflected in their free-form answers. Participants who had positive sentiments towards method A described it using terms such as ``understandable'', ``user friendly'', ``helpful'', and ``intuitive''.

We find that the following selection of the free-form answers represent the main reasons for which participants liked explanation method A: ``easy to distinguish the red (interesting) areas from the other areas which are not as important.'', ``Logical and a method that I have experienced with other examination modalities'', ``I find it visually easier to understand and it pinpoints exactly what it is reacting to so it is easy to double-check the data.'', ``It is simple and easy to understand which part of image should be focused on.''. Our interpretation of these and similar responses is that the saliency map produced by intrinsic explanation is visually intuitive and appealing, and therefore possibly also used in other applications the participants may be familiar with.

On the other hand, the positive assessment of explanation method A might depend strongly upon the model prediction being correct. One participant stated that method A is ``Easy to understand, the red part is mostly in the same spot as the lesion.'', suggesting that the assessment would have been different had the model not identified the polyp. Another participant's answer ``It helped me identify important areas.'' supports this notion.

Still, two participants wrote that ``I think it is the best one to help you focus on the area that the AI system has identified as a suspected polyp.'' and ``The red colour doesn't show up where the polyp is (\dots) so for someone used to identify polyps in colonoscopy [this doesn't indicate] that the program is able to really identify the polyp'', indicating that domain experts could use this explanation method to evaluate the machine learning model.

Among the responses giving method A a negative evaluation, one participant described the method as ``sensitive but less specific'' and another stated that it is ``Sometimes (\dots) a bit intense and difficult to interpret''. One participant complained about the accuracy of method A, stating that ``I prefer method A over B, but the poor accuracy of the method makes the explanation method (A) annoying rather than helpful.''. Finally, one participant stated that they ``Did not get an explanation'', which we interpret as alluding to the fact that highlighting which information goes into a decision is not sufficient to actually explain it.

\subsubsection{What do you think of explanation method B?}
As the Likert scale responses indicate, the study participants preferred the intrinsic explanation to extrinsic explanation. Based on similar free-form answers from several participants, describing method B as ``complicated'', ``confusing'', ``hard to interpret'', ``hard to understand'',  ``doesn't feel natural'', and ``harder to grasp visually'', we conclude that the extrinsic explanations functionality of highlighting in which direction each collection of image pixels drives the prediction, is counter-intuitive to domain experts not familiar with such a way to represent information. One participant also stated that ``It is difficult to understand where should we pay attention to. There are several green spots in one image.''. Even though this indicated that the extrinsic explanation is not preferred by domain experts, it does not mean that the method does not provide value.

Before participating in the study, the participants were given an introduction to each explanation method, but it seems that a short brief is not enough to become comfortable with the visualizations that the extrinsic explanation produces, as supported by one participant's statement ``The image is messy. I understand the method as explained, but the method makes no sense to me.'', and another's `` (\dots) hard for my brain to wrap itself around the red/green `type of data in agreement or not' paradigm used for this explanation''. Another stated that the method is ``Harder to understand - however after a while you get a hang of it'', indicating that more time spent contemplating the method or studying several examples could have a strong positive effect on the evaluation of the method by medical doctors. This notion is supported by participants stating that method B ``helps trust the system as it is based on the data'' and ``Makes more sense in terms of how data is trained''. The participants with positive sentiments towards explanation method B described it using terms such as ``Interesting'', ``intriguing'' and ``more specific''.

Further, it seems that the choice of color map as well as super pixel size could be adapted to better suit the end-users, as some participants stated that method B has ``not the best colours'', is ``confusing with the large amount of boxes not as pleasing to the eye'', and one suggests that ``colors should be opposite. Red for disease, green for healthy.''. The latter response indicates that the particular participant had misunderstood the method -- as the extrinsic method colouring indicates agreement with the model prediction; not the label -- and consequently that method B is not sufficiently intuitive, as recently discussed. One participant also voiced concern regarding aptness of this method for color blind people. We have not taken this aspect into account in our study, but stress that, in general, any visualization method should abide by the principles of universal design, including color blindness accessibility.

\section{Discussion}\label{section:discussion}
In general, the answers to the free form questions align with the responses to the Likert and multiple-choice questions. Most doctors prefer intrinsic-based explanations as they more easily align with their expectations in terms of spatial relevance and visual presentation. The participants found the intrinsic explanations more intuitive and user-friendly and that the visualizations more correctly aligned with their preconceived notions with regards to what they expected the model would react to. Specifically, some specifically state that they prefer the intrinsic explanation as it more accurately highlights the lesion. The problem here is that the explanations are not there to detect subjects in an image but rather to explain why a specific prediction was made. If the doctors expect the explanations to always align with the object in question, the explanations may hinder adoption and trust. As for the extrinsic explanations, several doctors were confused by the visualizations and found them hard to interpret, somewhat defeating the explanation in the first place. Some referenced the choice of colors and that the superpixels were not pleasing to look at. On the contrary, there seems to be a different level of understanding in terms of \gls{ai} knowledge among the participants, with some mentioning that they prefer the extrinsic explanation method due it providing more information about how the model was trained. Perhaps the superficial aspects of the explanation could be improved by involving potential end-users in the development process to tailor the explanations to fit their use-case and needs. By having a human-centered approach for generating explanations of AI systems, the explanations may be regarded as more useful for the end-users~\cite{Liao2020HumanCenteredXAI}. As for explanations in general, the study participants preferred that explanations were provided together with the model predictions (Figure~\ref{figure:histogram-explanation-or-not}), but what they regarded as the best explanation method varied between the participants. The human factor is important when developing model explanations~\cite{vellido2020XAIimportance}. What is regarded as a useful explanation by one person might not be so for another person. Consequently, subjective preferences might contribute to explain why the medical experts that participated in the study did not prefer the same type of model explanation.

\section{Conclusion}\label{section:conclusion}
This paper presents a study comparing intrinsic against extrinsic explanation methods from the perspective of medical doctors. The study was conducted using a \gls{gi} use-case involving explanations of a machine learning model used to predict polyps in images. Study participants were gathered from different parts of the globe to complete a survey consisting of model predictions accompanied by two explanation methods for ten different medical cases. Our results show that the intrinsic explanations are preferred. However, the free-form responses in our survey strongly suggest that the underlying reason for the doctors' preference of this method may be more superficial than actually understanding what information the different explanations convey. This suggests that a certain level of training or practice is required for the doctors to fully exploit the usefulness of \gls{ml} model explanations, although we might na{\"\i}vely expect that all image-based explanations are sufficiently intuitive to be useful without prior training. We highlight that any form of explanation targeted at non-technical end-users, such as doctors, must be developed with the end-user in mind, ideally also involving the end-user. This includes abiding by the principles of universal design, in order to accommodate specific needs. To conclude, medical doctors recognize the usefulness of visual explanations for deep learning-based computer-vision models, but limited understanding of functioning and the reasoning behind an explanation may lead to unwarranted judgements based on the wrong principles.

\bibliographystyle{elsarticle-num}
\bibliography{bibliography}

\begin{thebibliography}{10}
\expandafter\ifx\csname url\endcsname\relax
  \def\url#1{\texttt{#1}}\fi
\expandafter\ifx\csname urlprefix\endcsname\relax\def\urlprefix{URL }\fi
\expandafter\ifx\csname href\endcsname\relax
  \def\href#1#2{#2} \def\path#1{#1}\fi

\bibitem{Kelly2019}
C.~J. Kelly, A.~Karthikesalingam, M.~Suleyman, G.~Corrado, D.~King, {Key
  challenges for delivering clinical impact with artificial intelligence}, BMC
  Medicine 17~(1) (2019) 195.
\newblock \href {https://doi.org/10.1186/s12916-019-1426-2}
  {\path{doi:10.1186/s12916-019-1426-2}}.

\bibitem{Lundberg2017SHAP}
S.~M. Lundberg, S.-I. Lee, {A Unified Approach to Interpreting Model
  Predictions}, in: Proceedings of the 31st International Conference on Neural
  Information Processing Systems, Curran Associates, Inc., 2017, pp.
  4765--4774.

\bibitem{Selvaraju:2017:ICCV.2017.74}
R.~R. Selvaraju, M.~Cogswell, A.~Das, R.~Vedantam, D.~Parikh, D.~Batra,
  {Grad-CAM: Visual Explanations from Deep Networks via Gradient-Based
  Localization}, in: Proceedings of the IEEE International Conference on
  Computer Vision (ICCV), 2017, pp. 618--626.
\newblock \href {https://doi.org/10.1109/ICCV.2017.74}
  {\path{doi:10.1109/ICCV.2017.74}}.

\bibitem{vellido2020XAIimportance}
A.~Vellido, The importance of interpretability and visualization in machine
  learning for applications in medicine and health care, Neural computing and
  applications 32~(24) (2020) 18069--18083.
\newblock \href {https://doi.org/10.1007/s00521-019-04051-w}
  {\path{doi:10.1007/s00521-019-04051-w}}.

\bibitem{NEURIPS2020_2c29d89c}
J.~V. Jeyakumar, J.~Noor, Y.-H. Cheng, L.~Garcia, M.~Srivastava, How can i
  explain this to you? an empirical study of deep neural network explanation
  methods, in: H.~Larochelle, M.~Ranzato, R.~Hadsell, M.~F. Balcan, H.~Lin
  (Eds.), Advances in Neural Information Processing Systems, Vol.~33, Curran
  Associates, Inc., 2020, pp. 4211--4222.

\bibitem{Hoerter2020}
N.~Hoerter, S.~A. Gross, P.~S. Liang, {Artificial Intelligence and Polyp
  Detection}, Current treatment options in gastroenterology (jan 2020).
\newblock \href {https://doi.org/10.1007/s11938-020-00274-2}
  {\path{doi:10.1007/s11938-020-00274-2}}.

\bibitem{10.3389/fmed.2021.709347}
Y.-q. Song, X.-l. Mao, X.-b. Zhou, S.-q. He, Y.-h. Chen, L.-h. Zhang, S.-w. Xu,
  L.-l. Yan, S.-p. Tang, L.-p. Ye, S.-w. Li, Use of artificial intelligence to
  improve the quality control of gastrointestinal endoscopy, Frontiers in
  Medicine 8 (2021).
\newblock \href {https://doi.org/10.3389/fmed.2021.709347}
  {\path{doi:10.3389/fmed.2021.709347}}.

\bibitem{simonyan2014deep}
K.~Simonyan, A.~Vedaldi, A.~Zisserman, {Deep inside convolutional networks:
  Visualising image classification models and saliency maps}, in: In Workshop
  at International Conference on Learning Representations, 2014.

\bibitem{zhang2017mdnet}
Z.~Zhang, Y.~Xie, F.~Xing, M.~McGough, L.~Yang, {Mdnet: A semantically and
  visually interpretable medical image diagnosis network}, in: Proceedings of
  the IEEE conference on computer vision and pattern recognition, 2017, pp.
  6428--6436.

\bibitem{smilkov2017smoothgrad}
D.~Smilkov, N.~Thorat, B.~Kim, F.~Vi{\'e}gas, M.~Wattenberg, {Smoothgrad:
  removing noise by adding noise}, arXiv preprint arXiv:1706.03825 (2017).

\bibitem{sundararajan2017axiomatic}
M.~Sundararajan, A.~Taly, Q.~Yan, {Axiomatic attribution for deep networks},
  in: International Conference on Machine Learning, 2017, pp. 3319--3328.

\bibitem{bach2015pixel}
S.~Bach, A.~Binder, G.~Montavon, F.~Klauschen, K.-R. M{\"u}ller, W.~Samek, {On
  pixel-wise explanations for non-linear classifier decisions by layer-wise
  relevance propagation}, PloS one 10~(7) (2015) e0130140.

\bibitem{kindermans2017learning}
P.-J. Kindermans, K.~T. Sch{\"u}tt, M.~Alber, K.-R. M{\"u}ller, D.~Erhan,
  B.~Kim, S.~D{\"a}hne, {Learning how to explain neural networks: PatternNet
  and PatternAttribution} (2017).

\bibitem{montavon2017explaining}
G.~Montavon, S.~Lapuschkin, A.~Binder, W.~Samek, K.-R. M{\"u}ller, {Explaining
  nonlinear classification decisions with deep taylor decomposition}, Pattern
  Recognition 65 (2017) 211--222.

\bibitem{Adebayo2018SanityCheck}
J.~Adebayo, J.~Gilmer, M.~Muelly, I.~Goodfellow, M.~Hardt, B.~Kim, Sanity
  checks for saliency maps, in: S.~Bengio, H.~Wallach, H.~Larochelle,
  K.~Grauman, N.~Cesa-Bianchi, R.~Garnett (Eds.), Advances in Neural
  Information Processing Systems, Vol.~31, Curran Associates, Inc., 2018.

\bibitem{zeiler2014visualizing}
M.~D. Zeiler, R.~Fergus, {Visualizing and understanding convolutional
  networks}, in: European conference on computer vision, Springer, 2014, pp.
  818--833.

\bibitem{Achanta10slicsuperpixels}
R.~Achanta, A.~Shaji, K.~Smith, A.~Lucchi, P.~Fua, S.~S{\"u}sstrunk, {SLIC
  Superpixels} (2010).

\bibitem{10.1145/3193289}
K.~Pogorelov, K.~R. Randel, C.~Griwodz, S.~L. Eskeland, T.~de~Lange,
  D.~Johansen, C.~Spampinato, D.-T. Dang-Nguyen, M.~Lux, P.~T. Schmidt,
  M.~Riegler, P.~Halvorsen, Kvasir: A multi-class image dataset for computer
  aided gastrointestinal disease detection, in: Proceedings of the ACM on
  Multimedia Systems Conference (MMSYS), 2017, pp. 164--169.
\newblock \href {https://doi.org/10.1145/3083187.3083212}
  {\path{doi:10.1145/3083187.3083212}}.

\bibitem{He2015}
K.~He, X.~Zhang, S.~Ren, J.~Sun, Deep residual learning for image recognition,
  arXiv preprint arXiv:1512.03385 (2015).

\bibitem{kokhlikyan2020captum}
N.~Kokhlikyan, V.~Miglani, M.~Martin, E.~Wang, B.~Alsallakh, J.~Reynolds,
  A.~Melnikov, N.~Kliushkina, C.~Araya, S.~Yan, O.~Reblitz-Richardson, Captum:
  A unified and generic model interpretability library for pytorch (2020).

\bibitem{hallgren2012IRR}
K.~A. Hallgren, {Computing inter-rater reliability for observational data: an
  overview and tutorial}, Tutorials in quantitative methods for psychology
  8~(1) (2012) 23--34.

\bibitem{cicchetti1994guidelines}
D.~V. Cicchetti, Guidelines, criteria, and rules of thumb for evaluating normed
  and standardized assessment instruments in psychology., Psychological
  assessment 6~(4) (1994) 284--290.
\newblock \href {https://doi.org/10.1037/1040-3590.6.4.284}
  {\path{doi:10.1037/1040-3590.6.4.284}}.

\bibitem{GHASSEMI2021e745}
M.~Ghassemi, L.~Oakden-Rayner, A.~L. Beam, The false hope of current approaches
  to explainable artificial intelligence in health care, The Lancet Digital
  Health 3~(11) (2021) e745--e750.
\newblock \href {https://doi.org/10.1016/S2589-7500(21)00208-9}
  {\path{doi:10.1016/S2589-7500(21)00208-9}}.

\bibitem{Liao2020HumanCenteredXAI}
Q.~V. Liao, D.~Gruen, S.~Miller, Questioning the AI: Informing Design Practices
  for Explainable AI User Experiences, 2020, pp. 1--15.
\newblock \href {https://doi.org/10.1145/3313831.3376590}
  {\path{doi:10.1145/3313831.3376590}}.

\end{thebibliography}

\end{document}